\documentclass{pasj00}
\draft

\begin{document}
\SetRunningHead{Watanabe et al.}{Suzaku Observation of ZwCl0823.2+0425}
\Received{2010/08/20}
\Accepted{2010/12/27}

\title{Suzaku X-ray Follow-up Observation of Weak-lensing-detected Halos
       in the Field around ZwCl0823.2+0425}



%
 \author{%
   Eri \textsc{Watanabe}\altaffilmark{1},
   Motokazu \textsc{Takizawa}\altaffilmark{2},
   Kazuhiro \textsc{Nakazawa}\altaffilmark{3},
   Nobuhiro \textsc{Okabe}\altaffilmark{4,5},
   Madoka \textsc{Kawaharada}\altaffilmark{6},
   Arif \textsc{Babul}\altaffilmark{7}, 
   Alexis \textsc{Finoguenov}\altaffilmark{8,9},
   Graham P. \textsc{Smith}\altaffilmark{10},
   James E. \textsc{Taylor}\altaffilmark{11}
   }

 \altaffiltext{1}{School of Science and Engineering, Yamagata University, 
                  Kojirakawa-machi 1-4-12, Yamagata 990-8560}
 \altaffiltext{2}{Department of Physics, Yamagata University, Kojirakawa-machi
                  1-4-12, Yamagata 990-8560}
 \email{takizawa@sci.kj.yamagata-u.ac.jp}
 \altaffiltext{3}{Department of Physics, The University of Tokyo, 7-3-1 Hongo, 
                  Bunkyo-ku, Tokyo 113-0033}
 \altaffiltext{4}{Institute of Astronomy and Astrophysics, Academia Sinica, 
                  P.O.Box 23-141, Taipei 106, Taiwan}
 \altaffiltext{5}{Astronomical Institute, Tohoku University, Aramaki,
  Aoba-ku, Sendai, 980-8578}
 \altaffiltext{6}{Institute of Space and Astronautical Science/JAXA, 3-1-1 Yoshinodai,
                  Chuo-ku, Sagamihara, Kanagawa 252-5210}
 \altaffiltext{7}{Department of Physics and Astronomy, University of Victoria, Victoria, BC, V8P 1A1, Canada}
 \altaffiltext{8}{Max-Planck-Institut f\"{u}r extraterrestrische Physik, Giessenbachstra\ss e, 
                  85748 Garching, Germany}
 \altaffiltext{9}{University of Maryland, Baltimore County, 1000 Hilltop Circle, Baltimore, MD 21250, USA}
 \altaffiltext{10}{School of Physics and Astronomy, University of Birmingham, Edgbaston, 
                  Birmingham, B15 2TT, UK}
 \altaffiltext{11}{Department of Physics and Astronomy, University of Waterloo, 200 University Avenue West, 
                   Waterloo, ON N2L 3G1, Canada}
\KeyWords{galaxies: clusters: individual (ZwCl0823.2+0425) --- X-rays: galaxies: 
          clusters --- gravitational lensing} 

\maketitle

\begin{abstract}
  We present the results of Suzaku X-ray follow-up observation of weak-lensing-detected halos in 
  the field around galaxy cluster ZwCl0823.2+0425. We clearly detected X-ray emission
  associated with most of these halos and determined their detailed physical parameters 
  such as X-ray luminosity, temperature, and metal abundance, for the first time.
  We find that the X-ray luminosity - temperature relation for these halos agrees with
  former typical results. With mass determined from the weak gravitational lensing data, 
  the mass-temperature relation for them is also investigated and found to be 
  consistent with the prediction from a simple self-similar model
  and results of the previous studies with both lensing and X-ray data.
  We would like to emphasize that the self-similar scaling relation of mass and temperature 
  is shown here for the first time using a weak-lensing selected sample, whereas previous
  studies of the mass scaling relation used X-ray-selected samples of clusters.
  Therefore, our study demonstrates importance of X-ray follow-up
  observations of shear-selected clusters, and shows that
  a joint X-ray and lensing analysis will be crucial for clusters discovered by the
  forthcoming weak-lensing surveys, such as the one planned with Subaru/Hyper-Suprime-Cam.
\end{abstract}

\section{Introduction}\label{s:INT}
In the standard theory of structure formation in the cold dark matter (CDM)
universe, larger structures such as dark halos corresponding to rich clusters of galaxies 
form through absorption of smaller dark halos and accretion of surrounding matter,
which preferably occur along filaments associated with large scale structures in the universe.
Although baryonic components such as diffuse hot gas and galaxies are visible
in X-ray and optical bands, dark matter, which is a dominant component in mass, 
cannot be directly seen in any kind of radiation, 
and we recognized its presence only indirectly through dynamical properties of galaxies and hot gas.

In recent years, however, this situation 
has been changed by the development of gravitational lensing techniques,
which enable
us to reveal matter distribution in the universe directly.
We do not require any assumptions of the dynamical status and
mass-luminosity ratio of the system to get the mass distribution \citep{Schn06}. 
Whereas strong lensing technique is essentially model dependent, 
weak gravitational lensing enables us to obtain
mass distribution in the plane of the sky without any geometrical assumptions
and measures directly galaxy cluster mass distribution
out to the virial radius \citep{Kais93, Bart01}.
However, it is true that we cannot get detailed properties of baryonic components
associated with dark matter halos by gravitational lensing alone.
Therefore, joint studies of X-ray, optical and weak lensing datasets are very useful 
and provide us with important information of the interplay between baryonic and dark matter 
\citep{Okab08, Kawa10, Umet10}.

In the very near future, large surveys in various wavebands
(e.g.\ Subaru/Hyper-Suprime-Cam, eROSITA, SPT, and ACT) 
will provide us with huge datasets of galaxy clusters.
In such future surveys, the weak gravitational lensing technique 
is one of the most powerful methods to find galaxy clusters, irrespective of
the physical status of the baryonic components \citep{Witt01, Miya02b, Miya07}. 
Its detection efficiency of clusters/groups is a function of
masses as well as the geometry of the universe. 
A typical procedure is as follows; the lensing convergence field is reconstructed from galaxy's
ellipticities convolved with a suitable filter \citep{Kais93, Seit01}, and lensing signals above
the noise of intrinsic ellipticities are identified as mass
structures \citep{Witt01, Miya02b, Miya07}.
However, there are still some uncertainties in this method such as 
contaminations of member galaxies in shear catalogs, spurious peaks
induced by intrinsic ellipticities, and the projection effects due to
superpositions of structures at different redshifts \citep{Miya02b,Hama09,Okab10a}. 
Owing to not only these uncertainties characteristic of the weak lensing surveys but also
the systematic difference of detection efficiency among weak lensing, X-ray,
and Sunyaev-Zel'dovich effect (SZE) surveys, it is still unclear whether 
halo candidates detected by weak lensing always contain hot baryons. 
Thus, X-ray and/or SZE follow-up observations of halo candidates discovered by weak
lensing analysis are very important as a pilot study before forthcoming large surveys 
in order to understand systematic differences among different wavelength surveys 
and the correlations of physical properties between hot baryons and total mass.
In particular, a study of the scaling relations between lensing mass and
baryonic observables is of prime importance for the
cluster cosmology as well as understanding cluster baryonic
evolution, such as radiative cooling, energy feedback, and cluster
mergers \citep{Stan10,Vikh09a,Vikh09b,Okab10c}.

We conducted Suzaku X-ray follow-up observation of weak-lensing-detected
halo candidates.
X-ray Imaging Spectrometer (XIS) aboard Suzaku satellite \citep{Mits07} is
suitable for observing low surface brightness diffuse sources
thanks to its low and stable background \citep{Koya07}. 
Indeed, physical properties of the intracluster medium (ICM)
around the virial radius, where X-ray emission is too faint for Chandra and XMM-Newton to 
get meaningful results, 
have been investigated for several clusters \citep{Fuji08, Baut09, Geor09, Reip09, Kawa10, Hoshi10}. 
This feature is also very useful to investigate diffuse X-ray emission associated 
with mass structures, even less massive objects, found by weak lensing surveys.

As a pilot study for the above-mentioned purpose, 
we select a field around galaxy cluster ZwCl0823.2+0425. Several dark
halos are found around the cluster by weak lensing study \citep{Okab10b} as a collaboration of
the Local Cluster Substructure Survey (LoCuSS; PI: Graham P. Smith).
Figure \ref{fig1} shows optical image of the field overlaid with the mass contours
derived from the weak lensing analysis of \citet{Okab10b}. Besides the central halo corresponding to the
ZwCl0823.2+0425, four significant mass clumps are seen in the north, north-west, north-east, 
and south-east of it. We refer them as ZwCl0823.2+0425 (or C), N, NW, NE, and SE hereafter.
The mass peak corresponding to Abell 664 is also found near the boundary.
Two major red-sequence galaxy populations are found in this field, whose spatial distributions of 
both optical luminosity and number density are pretty similar to that of mass \citep{Okab10b}. 
The galaxies in the red-sequence at the lower redshift are apparently associated with 
the C and NW halos. On the other hand, those in the higher redshift red-sequence seem to be
related with the N and NE halos. 
The SDSS spectroscopic data are available for a few galaxies located in each halo center.
We found that the mean redshift of galaxies associated with the C and NW
halos is $z=0.2248$, and that of N and NE halos is $z=0.472$. 
The SE halo is likely to be associated with a bright galaxy at $z=0.103$. However, a few galaxies
in the background possibly contribute the lensing signal. This possible
projection effect will be discussed in section \ref{s:D}.
Although ZwCl0823.2+0425 is recognized as an X-ray source with the flux
of $1.8 \times 10^{-12}$ erg s$^{-1}$ cm$^{-2}$ in Bright Source Catalog of ROSAT All Sky Survey
\citep{Ebel98, Ebel00, Bohr04},
its detailed spectral properties are not known. Moreover, X-ray emission associated with 
the other halos are not clearly detected. Therefore, deeper X-ray observation for this field
is highly desirable to explore physical properties of the hot gas associated with the halos.

In this paper, we present Suzaku X-ray follow-up observation of the field around
ZwCl0823.2+0425 to investigate the physical properties of the hot gas in the
dark matter halos found via gravitational lensing analysis. 
The rest of this paper is organized as follows. In section \ref{s:OD} we
describe the observation and data reduction. In section \ref{s:SA} we
present spectral analysis results. In section \ref{s:IS} we describe the imaging simulation
to check the contamination of each halo's spectrum from the others. In section \ref{s:D} we discuss the
results and their implications. In section \ref{s:C} we summarize the results.
Canonical cosmological parameters of $H_0 = 70$ Mpc$^{-1}$ km s$^{-1}$,
$\Omega_0=0.27$, and $\Lambda_0=0.73$ are used in this paper. Unless
otherwise stated, all uncertainties are given at the 90\% confidence level.

\section{Observation and Data Reduction}\label{s:OD}
\begin{figure}
  \begin{center}
    \FigureFile(80mm,80mm){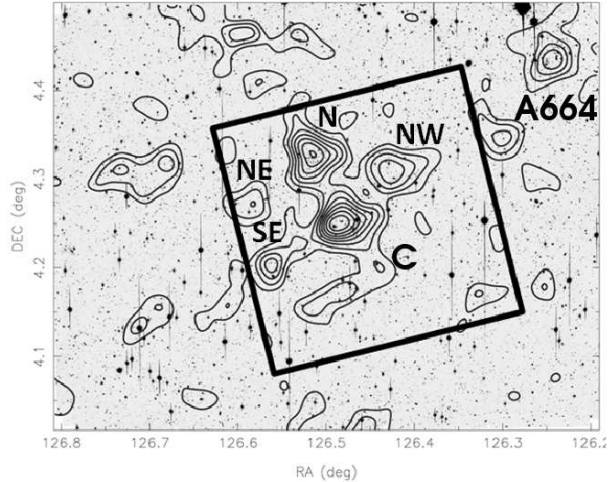}
  \end{center}
  \caption{Optical image of the field around ZwCl0823.2+0425 overlaid with the mass contours
           derived from the weak lensing analysis. The field of view of Suzaku XIS is also 
           shown in a black square. In addition to the central mass peak corresponding 
           to ZwCl0823.2+0425, four mass peaks are seen around it. Hereafter, 
           we refer them as C, N, NW, NE, and SE, as shown in this figure. The mass
           peak corresponding to Abell 664 is also found outside of the XIS field of view.
           The contours are spaced in a unit of $1\sigma$ reconstruction error, $\delta \kappa=0.02243$.
                }\label{fig1}
\end{figure}

We observed the field around ZwCl0823.2+0425 with Suzaku on 2008 May 17-18. The field of view (FOV)
of Suzaku XIS is shown in a Subaru optical image overlaid with the mass contours derived from
the weak lensing analysis \citep{Okab10b} in figure \ref{fig1}. 
The observation was performed at XIS nominal 
pointing. The XIS was operated in the normal full-frame clocking mode. The edit mode was $3 \times 3$
and $5 \times 5$, and we used combined data of both modes. The spaced-row charge injection was adopted
for XIS. All data were processed with Suzaku pipeline processing, version 2.2. 
We employed calibration data files (20090925).

\begin{figure}
  \begin{center}
    \FigureFile(80mm,80mm){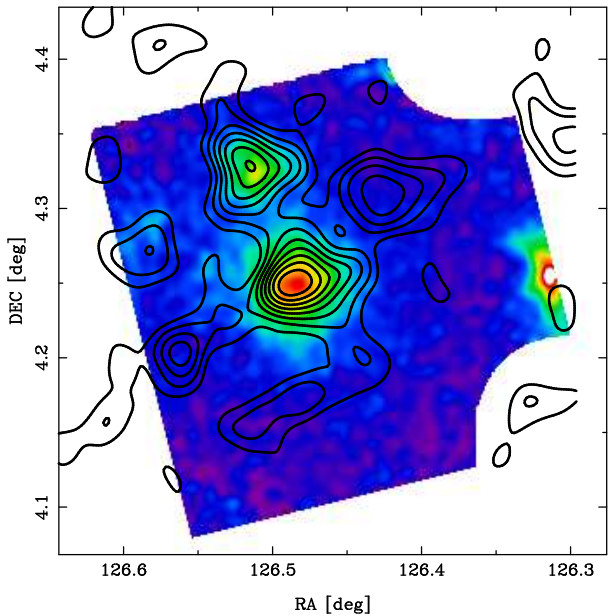}
  \end{center}
  \caption{XIS image of the field around ZwCl0823.2+0425 in the 0.5-8.0 keV band
           combined from XIS FI CCD images, overlaid with the mass contours. 
           The image was corrected for exposure and
           vignetting effects after subtracting NXB, and smoothed by a Gaussian kernel
           with $\sigma=0'17$.}\label{fig2}
\end{figure}

The XIS data were processed through standard criteria as follows. Events with a GRADE of 
0, 2, 3, 4, 6 and STATUS with 0:524287 were extracted. We excluded data obtained at the South Atlantic
Anomaly (SAA), within 436s after the passage of SAA, and at low elevation angles from an Earth rim of 
$<5^{\circ}$ and a Sun-lit Earth rim of $<20^{\circ}$. As a result, effective exposure time was 41.3 ks.
Non X-ray background (NXB) spectra and images of XIS were generated using the ftool ``xisnxbgen'' 
\citep{Tawa08}. 
Figure \ref{fig2} represents an 0.5-8.0 keV XIS image combined from those of the
front illuminated (FI) CCDs (XIS0, XIS3), overlaid with the mass contours. 
The image was corrected for exposure and vignetting effects
after subtracting NXB, and smoothed by a Gaussian kernel with
$\sigma=0'.17$.
Enhanced X-ray emission from the ZwCl0823.2+0425, N and NE halo regions are
clearly apparent in figure \ref{fig2}. 

\section{Spectral Analysis}\label{s:SA}
For a spectral analysis of the XIS data, redistribution matrix files (RMFs) 
were generated using the ftool ``xisrmfgen''.
In addition, ancillary response files (ARFs), which describe the response of X-Ray Telescope
aboard Suzaku and the amount of the XIS optical blocking filters contamination, were generated with
the ftool ``xissimarfgen'' \citep{Ishi07}.
XIS spectra for each sensor (XIS0, XIS1, XIS3) were fitted simultaneously. 
For the spectral fitting, we used the energy band of
0.5 -- 10.0 keV and 0.5 -- 8.0 keV for FI CCDs (XIS0 and XIS3) and 
back illuminated (BI) one (XIS1), respectively.
In the spectral fitting with Galactic absorption, we assumed $N_H=3.18 \times 10^{20}$ cm$^{-2}$
\citep{Dick90}.
The CXB level was estimated in the same way as \citet{Naka09} from the Lockman hole
observation (Suzaku observation ID, 101002010). The detailed procedure is described
in appendix 1 of \citet{Naka09}. We defined the photon flux model as 
$N(E)=9.19 \times 10^{-4} \times E^{-1.4}$ in photons cm$^{-2}$
s$^{-1}$ keV$^{-1}$ FOV$^{-1}$, where $E$ is the photon energy in keV.

\subsection{The Background Model}
\begin{figure}
  \begin{center}
    \FigureFile(80mm,80mm){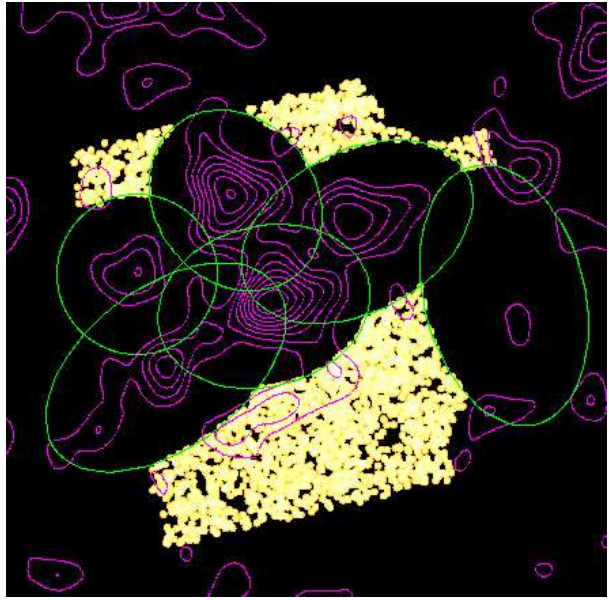}
  \end{center}
  \caption{A Region used to determine the background model (yellow field). 
           We eliminated regions corresponding to each halo and the foreground bright star near 
           the west edge of the XIS FOV (green circles and ellipses).}\label{fig3}
\end{figure}

Before entering the spectral analysis of each halo, we need to construct a background model
from astrophysical origin in addition to NXB. As shown in figure \ref{fig3}, we eliminated
regions corresponding to each halo and the foreground bright star near the west edge of the XIS FOV,
the rest of which was used for spectral fit to determine the background model.
For the analysis of the background model, uniform emission over a circular region with 20' radius
was used as an input image to generate an ARF. The Cosmic X-ray background (CXB), 
the Galactic halo's hot gas (GH), and the local hot bubble (LHB) were considered 
as the background components. Then, the spectrum of the background region was fitted by a model as follows,
\begin{eqnarray}
   apec1 + wabs \times ( apec2 + apec3 + powerlaw )
\end{eqnarray}
where $apec1$, $apec2+apec3$, and $powerlaw$ represent the LHB, GH, and CXB, respectively.
The temperature of the LHB was fixed to be 0.08 keV. The metal abundance of the both LHB and GH
were also fixed to be solar.

\begin{figure}
  \begin{center}
    \FigureFile(80mm,80mm){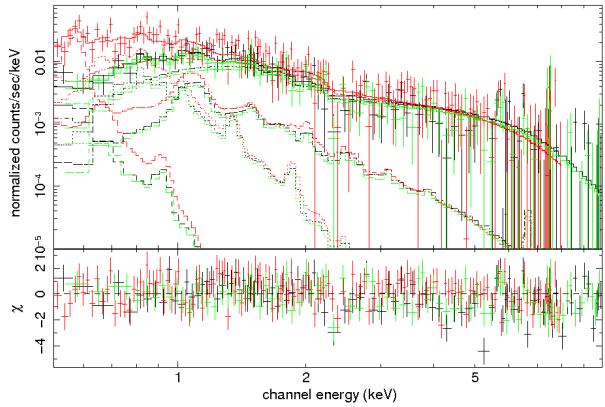}
  \end{center}
  \caption{XIS spectrum of the region shown in figure \ref{fig3} fitted with the background model described
           in the text. The black, red, and green crosses show the spectrum of XIS0, XIS1, and XIS3,
           respectively. The total and each component of the best fit model spectra are also plotted 
           as solid and dashed histograms, respectively.}\label{fig4}
\end{figure}

The spectrum of the background region fitted with the above-mentioned model is shown in figure 
\ref{fig4}, where the black, red, and green crosses show the spectrum of XIS0, XIS1, and XIS3,
respectively. The total and each component of the best fit model spectra are also plotted as solid and dashed 
histograms, respectively.
The detailed results of the fit are summarized in table \ref{tab:spec_back}.
At first glance, the higher temperature component of GH may seem to be too high. We tried to fit the
background region spectrum by a model with a single temperature GH. However, the results were not so good
and we had some residuals around $\sim 2$ keV in the data. This might be because our spectral modeling 
is too simple for the data. However, our main purpose here is to construct a plausible background 
spectrum model to investigate the halos' hot gas. Thus, we do not pursue this issue in more detail.
\begin{table}
  \caption{Best fit parameters for the XIS spectrum of the background region.}
  \label{tab:spec_back}
  \begin{center}
    \begin{tabular}{ll}
      \hline
               &  The background model \\
      \hline
      $kT_{\rm LHB}$\footnotemark[$*$]          &  0.08(fixed)  \\
      $N_{\rm LHB}$\footnotemark[$\dagger$]     &  $1.19^{+0.35}_{-0.40} \times 10^{-2}$  \\
      $kT_{\rm GH, low}$\footnotemark[$*$]      &  $0.34^{+0.24}_{-0.0542}$    \\
      $N_{\rm GH, low}$\footnotemark[$\dagger$] &  $5.76^{+1.90}_{-2.24} \times 10^{-4}$  \\ 
      $kT_{\rm GH, high}$\footnotemark[$*$]     &  $1.71^{+0.45}_{-0.38}$    \\
      $N_{\rm GH, high}$\footnotemark[$\dagger$] &  $5.18^{+1.73}_{-2.11} \times 10^{-4}$  \\ 
      $\Gamma_{\rm PL}$\footnotemark[$\ddagger$] &   1.4(fixed)                          \\
      $N_{\rm PL}$\footnotemark[$\S$]      &  $9.19 \times 10^{-4}$(fixed)     \\
      $\chi^2/$d.o.f.             &  403.5/336                           \\

      \hline \\
   \multicolumn{2}{@{}l@{}}{\hbox to 0pt{\parbox{180mm}{\footnotesize
       \footnotemark[$*$] Temperature of the each component in keV.
       \par\noindent
       \footnotemark[$\dagger$] Normalization in the $apec$ code for each component.
       \par\noindent
       \footnotemark[$\ddagger$] Photon index of the power-law component.
       \par\noindent
       \footnotemark[$\S$] Normalization in the power-law component.
     }\hss}}
    \end{tabular}
  \end{center}
\end{table}

\subsection{Hot Gas in Each Halo} \label{S:gas}
\begin{figure}
  \begin{center}
    \FigureFile(80mm,80mm){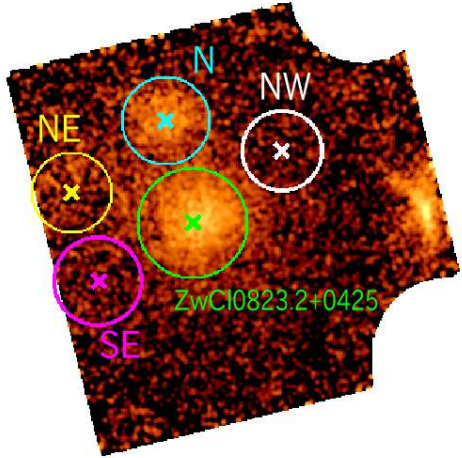}
  \end{center}
  \caption{Regions used in the spectral analysis of each halo.}\label{fig5}
\end{figure}

We performed spectral analysis for each halo using the background model in
the previous subsection. Figure \ref{fig5} shows regions used in the
analysis of each halos, where the center of each region was determined
through the each halo's mass peak in the weak lensing data of
\citet{Okab10b}. The radii of the regions are 2.5', 2.0', 1.8', 1.8', and
2.0' for C, N, NE, NW, and SE, respectively. 
It is true that these radii should be as large as the expected virial radii of the corresponding halo.
However, because each halo partially overlaps with one another and the spatial resolution of Suzaku 
is moderate, the radii have to be small enough to reduce contamination from the others.
No core removal was made in the spectral extraction.
We adopted model images of the $\beta$-model profile with core radius 0.2 Mpc and $\beta=0.66$ 
at each halo's redshift as an input image to generate ARFs, considering 
that Suzaku does not have spatial resolution good enough to resolve inner structures of each halo
and that there is no X-ray image of this field deep enough for our purpose by other instruments. 

Systematic errors of both CXB and NXB were taken into account in the
following analysis. It is well-known that the CXB fluctuations can be modeled as 
$\sigma_{\rm CXB}/I_{\rm CXB} \propto \Omega_e^{-0.5} S_c^{0.25}$, 
where $\Omega_e$ and $S_c$ are the effective solid angle and upper cutoff flux of a point source, 
respectively. From the HEAO-1 A2 results, $\sigma_{\rm CXB}/I_{\rm CXB}=2.8$\% 
with $\Omega_e=15.8$ deg$^2$ and $S_c = 8 \times 10^{-11}$ erg s$^{-1}$ cm$^{-2}$ \citep{Shaf83}. 
We adopted upper cutoff flux $S_c \sim 1.0 \times 10^{-14}$ erg s$^{-1}$ cm$^{-2}$ taking into account 
the fact that point sources with similar flux were clearly detected in the Lockman hole observation 
(Suzaku observation ID, 101002010).
Thus, the CXB fluctuations for C, N, NE, NW are expected to be 26, 33, 36, and 36 \% 
at the 90\% confidence level. 
\citet{Tawa08} reported that reproducibility of NXB was 6.0\% and 12.5\%
for XIS FI and BI at the 90\% confidence level, respectively.
We did not take account of systematic errors of LHB and GH, considering that \citet{Yosh09}
reported that prominent spatial fluctuations in a scale smaller than that
of XIS FOV were hardly seen in LHB and GH.
In the spectrum analysis, NXB and CXB components were fluctuated at the 90 \%
confidence level of the systematic uncertainty mentioned above, which caused the changes of 
the best-fit parameters in the fits and gave us the systematic errors.

\begin{figure}
  \begin{center}
    \FigureFile(80mm,80mm){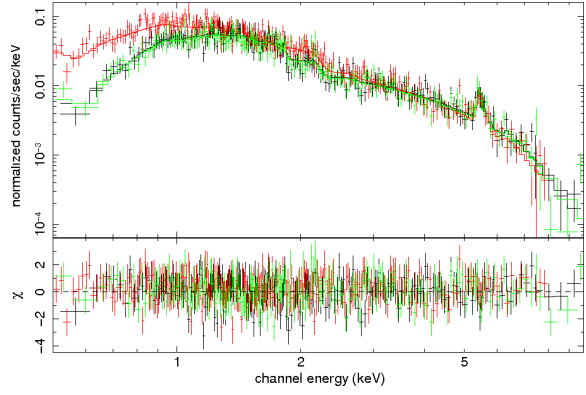}
  \end{center}
  \caption{XIS spectrum of the ZwCl0823.2+0425 fitted with the model described in the text.
           The black, red, and green crosses show the spectrum of XIS0, XIS1, and XIS3,
           respectively. The best fit model spectra are plotted as solid histograms.}\label{fig6}
\end{figure}

\begin{figure}
  \begin{center}
    \FigureFile(80mm,80mm){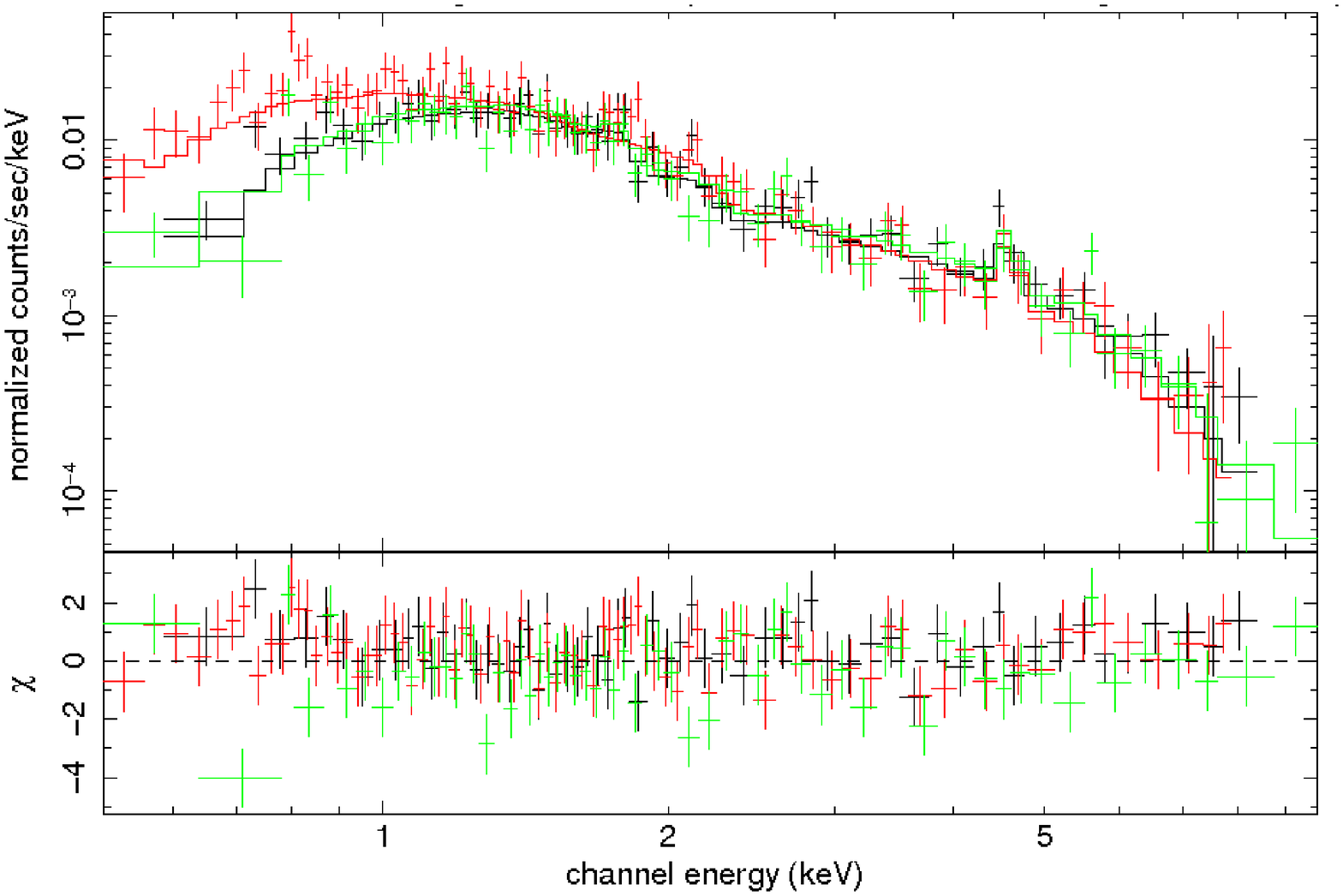}
  \end{center}
  \caption{Same as figure \ref{fig6}, but for the N halo.}\label{fig7}
\end{figure}

\begin{figure}
  \begin{center}
    \FigureFile(80mm,80mm){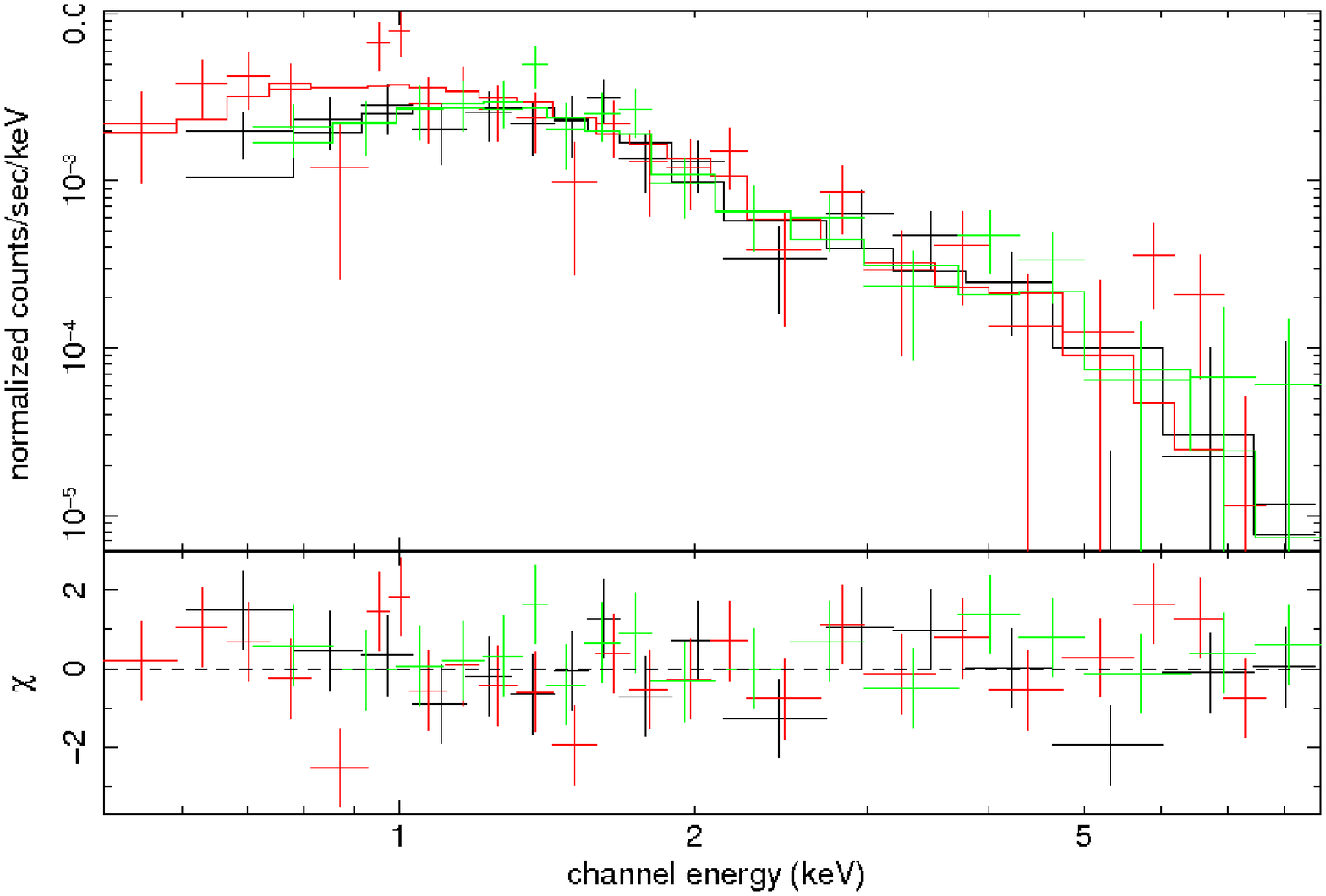}
  \end{center}
  \caption{Same as figure \ref{fig6}, but for the NE halo.}\label{fig8}
\end{figure}

\begin{figure}
  \begin{center}
    \FigureFile(80mm,80mm){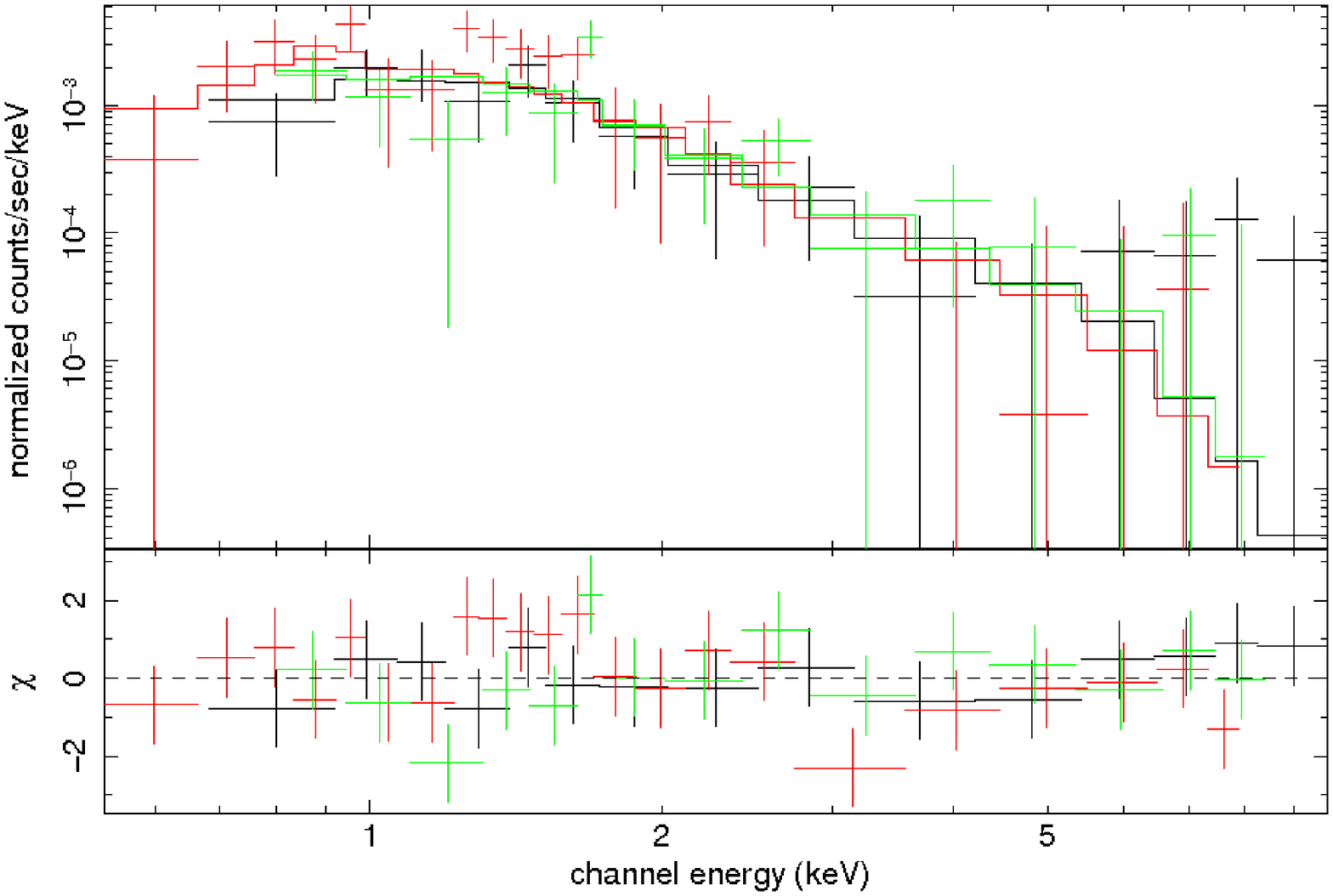}
  \end{center}
  \caption{Same as figure \ref{fig6}, but for the NW halo.}\label{fig9}
\end{figure}

\begin{figure}
  \begin{center}
    \FigureFile(80mm,80mm){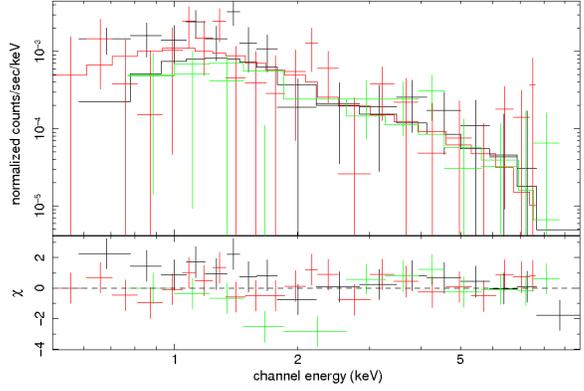}
  \end{center}
  \caption{Same as figure \ref{fig6}, but for the SE halo. Unfortunately, however, the statics of 
           the SE spectrum is not good enough to obtain meaningful fitting results.}\label{fig10}
\end{figure}

Each halo's spectrum was fitted by the photoabsorbed single temperature apec model 
($wabs \times apec$) with the background model described in the previous subsection. 
The redshift of each halo 
was fixed to be a value obtained from the SDSS data. We fixed the metal abundance for NE, NW, and SW
to be 0.3 solar because we do not have enough photons for these regions to determine it.
Figure \ref{fig6}, \ref{fig7}, \ref{fig8}, \ref{fig9}, and \ref{fig10} represent the XIS spectrum 
fitted with the above-mentioned model of C, N, NE, NW, and SE, respectively.
In each figure, the black, red, and green crosses show the spectrum of XIS0, XIS1, and XIS3,
respectively. The best fit model spectra are also plotted as solid histograms. In general, 
all data except for SE are fitted well by the adopted model.
Unfortunately, however, the statics of the SE spectrum is not good enough to obtain meaningful 
fitting results. This is consistent with the fact that SE lensing signal in the mass map is due 
to the projection effect, which is mainly associated with a galaxy at low redshift $z\sim0.1$,
as described in section \ref{s:D}.
The detailed fitting results are presented in table \ref{tab:spec_halos}.
The luminosity in $0.5-10.0$ keV of each halo is calculated from the fitting
results and summarized in table \ref{tab:lumi_halos}.
\begin{table}
  \caption{Best fit parameters for the XIS spectrum of each halo region presented in figure
           \ref{fig6}, \ref{fig7}, \ref{fig8}, and \ref{fig9}.}
  \label{tab:spec_halos}
  \begin{center}
    \begin{tabular}{lllll}
      \hline
                         &  C & N & NE & NW \\
      \hline
 $kT$\footnotemark[$*$]    & $4.83^{+0.22+0.09}_{-0.21-0.09}$ & $6.35^{+0.75+0.24}_{-0.65-0.25}$ & 
                       $3.79^{+1.26+0.56}_{-0.90-0.66}$  & $2.04^{+0.75+0.57}_{-0.44-0.42}$  \\
 $Z$\footnotemark[$\dagger$]  & $0.33^{+0.07+0.00}_{-0.07-0.00}$ & $0.30^{+0.15+0.01}_{-0.14-0.01}$ &  0.3(fixed) &  0.3(fixed) \\
 $z$\footnotemark[$\ddagger$] &  0.2248(fixed)                & 0.472(fixed)  &   0.472(fixed)  &  0.2248(fixed) \\
 $N$\footnotemark[$\S$]     &  $3.44^{+0.08+0.04}_{-0.09-0.04} \times 10^{-3}$  & $1.51^{+0.07+0.04}_{-0.07-0.04} \times 10^{-3}$ &
                    $4.04^{+0.50+0.22}_{-0.56-0.35} \times 10^{-4}$  & $1.73^{+0.23+0.37}_{-0.36-0.46} \times 10^{-4}$ \\
 $\chi^2/$d.o.f. &  598.0/642  &  204.1/210 &   48.9/60.0  &  40.8/52 \\
      \hline \\
   \multicolumn{5}{@{}l@{}}{\hbox to 0pt{\parbox{180mm}{\footnotesize
       \footnotemark[$*$] Temperature of the each halo in keV. Errors are 90\% statistical and systematic, 
                          respectively. The latter is evaluated by shifting the NXB and CXB within their 
                          systematic errors.
       \par\noindent
       \footnotemark[$\dagger$] Metal abundance.
       \par\noindent
       \footnotemark[$\ddagger$] Redshift obtained from the SDSS data for each halo.
       \par\noindent
       \footnotemark[$\S$] Normalization in the apec code.
     }\hss}}
    \end{tabular}
  \end{center}
\end{table}
\begin{table}
  \caption{Luminosity in 0.5-10.0 keV of each halo derived from the fitting results listed 
           in table \ref{tab:spec_halos} with statistical and systematic errors.}
  \label{tab:lumi_halos}
  \begin{center}
    \begin{tabular}{lllll}
      \hline
               & 
 C & N & NE & NW \\
      \hline
      Luminosity ($10^{44}$ erg s$^{-1}$)  & $4.76^{+0.11+0.08}_{-0.14-0.08}$   & $8.82^{+0.44+0.34}_{-0.57-0.31}$  & 
                                            $1.89^{+0.21+0.25}_{-0.25-0.33}$   & $0.15^{+0.05+0.06}_{-0.05-0.05}$ \\
      \hline
    \end{tabular}
  \end{center}
\end{table}

We used model images to generate ARFs because Suzaku has 
only moderate spatial resolution ($\sim 2'$), and because we do not have appropriate X-ray images 
with better spatial resolution by other instrument for this field. However, it should be noted that
an effective area for diffuse sources depends on their spatial distribution. Therefore,
absolute values of the apec code's normalizations in table \ref{tab:spec_halos} and 
fluxes listed in table \ref{tab:lumi_halos} could change if the adopted model images change.
To evaluate this, we generated ARFs with a point source and $\beta$-model images with different
core radii ($r_{\rm c}=0.1$ and $0.3$ Mpc). 
We confirmed that this changed the fluxes by $\sim 50$ \% at most. Thus, we believe that
uncertainty due to image modeling for ARF generation is at a few tens of percent level.

Some of the halos might have central cool cores that are not resolved in our observation.
This means that the halos' temperature and metal abundance 
are possibly underestimated and overestimated, respectively. 
Furthermore, unresolved point sources with harder spectrum
such as AGNs could be contaminated, which results in overestimation of both the ICM temperature and metal
abundance. If these kinds of unresolved contamination sources are dominant,  
the observed metal abundance could be higher than that of a typical cluster without a cooling core.
However, such a trend is not seen at least in the central cluster and N halo. 
Thus, we safely conclude that the effects are quite limited at least for these halos.

\section{Imaging Simulation} \label{s:IS}
\begin{figure}
  \begin{center}
    \FigureFile(80mm,80mm){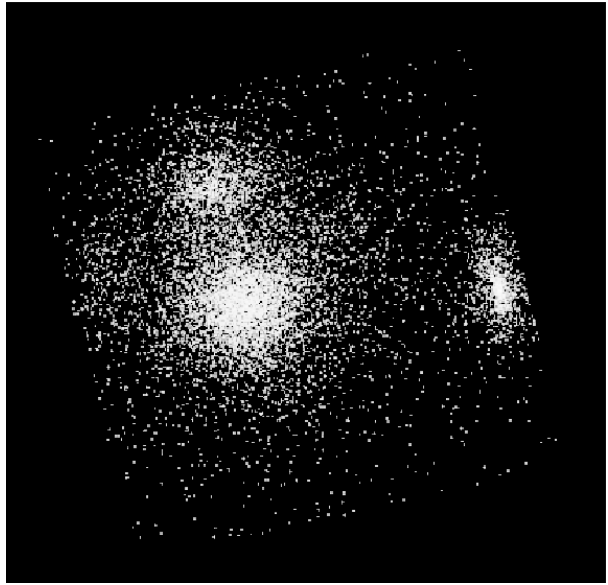}
  \end{center}
  \caption{Simulated image of XIS0.}\label{fig11}
\end{figure}

Because each halo might partially overlap with one another, we used regions with radii of only
$\sim 2'$ as shown in figure \ref{fig5}. 
Still, each halo's spectrum could be contaminated by the other halos and point source
near the west edge of XIS FOV. 
We simulated X-ray emission from them with the ftool ``xissim'' 
in order to estimate contamination to each halo region from the others. 
We placed five $\beta$-model X-ray halos with $r_{\rm c}=0.2$ Mpc at the position 
of C, N, NE, NW, and SE,
and one point source at the position of the bright star. As a spectrum model, we used the best-fit 
models listed in table \ref{tab:spec_halos} for C, N, NE, and NW. 
We also fit the SE and point source 
in a similar way, and the best-fit results were used as a spectrum model for the image simulation.
The resultant simulated image of XIS0 is shown in figure \ref{fig11}.

We checked contamination to each halo region from the other halos and point sources. 
We confirmed that only 4 \% of photons detected in the C region was from the others. 
For N, 16 \% of photons were from the others, most of which was originated from C. 
32 \% of photons detected in NE regions were from the others, 
about two thirds and one third of which were from C and N, respectively.
As for NW, 56 \% of photons detected in the region was from others, 
about three quarters and one quarter of which were originated from C and N, respectively.
Even in case of the faintest NW halo, contamination from the others alone cannot account for the detected
X-ray emission. 
It should be noted that relatively faint halos' (NE and NW) spectrum are significantly contaminated
by those of the brighter ones (C and N) with higher temperature, which means that the temperatures of
NE and NW are probably overestimated.
This effect is not corrected in the spectrum analysis because our imaging simulation is just aimed at
confirming that the fainter halo's emission cannot be accounted for only
the contamination from the brighter ones and that the hot gas associated with these fainter halos is surely 
detected.

\section{Discussion}\label{s:D}
\begin{figure}
  \begin{center}
    \FigureFile(80mm,80mm){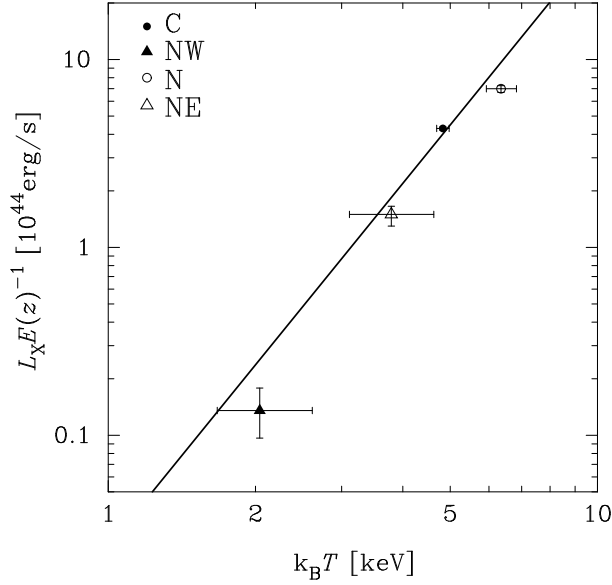}
  \end{center}
  \caption{X-ray luminosity - temperature relation for the halos. Errors are converted 
           into the one-sigma level. The best-fit power-law function (equation (\ref{eq:lxkt}))
           is also overlaid.}\label{fig12}
\end{figure}

First, we investigate the X-ray luminosity-temperature ($L_X-k_{\rm B}T$) relation for 
the halos.  Figure \ref{fig12} shows the $L_X-k_{\rm B}T$ relation
obtained from our results, where the errors are converted into at the one-sigma level.
The best-fit power-law function is 
\begin{eqnarray}
  \log L_X (10^{44}{\rm erg}\ \ {\rm s}^{-1}) E(z)^{-1} =  -1.59^{+0.47}_{-0.38} 
             + 3.21^{+0.52}_{-0.69} \log k_{\rm B}T ({\rm keV}),
  \label{eq:lxkt}
\end{eqnarray}
where 
\begin{eqnarray}
        E(z) = \sqrt{\Omega_0 (1+z)^3 + \Lambda_0}.
\end{eqnarray}
Although errors are large because of relatively poor statistics, our results are consistent 
with the former typical ones,
which show $L_X E(z)^{-1} \propto T^{2.5-3.0}$ (e.g., \cite{Ikeb02, Bran07, Ohar07}).

Second, let us investigate the mass-temperature ($M-k_{\rm B}T$) relation.
Here, we made a secure selection of background shear catalog in a color magnitude plane
in order to avoid contaminations of member galaxies associated
with each clump \citep{Umet08,Okab10b}. As described in section \ref{s:INT}, 
there are two red-sequences at $z\sim0.22$ and $\sim0.47$. Thus, 
we conservatively cut the broad range of the color. 
The resultant number density of the background galaxies decreased to
one-fourth of the original shear catalog ($n_g\simeq7.6~{\rm [arcmin^{-2}]}$).
We calibrated the mean redshifts based on the COSMOS photometric
redshift catalog \citep{Ilbe09}.  
The existence of halos at various redshifts makes it difficult to conduct 
one-dimensional tangential distortion analysis in order to measure each halo
mass. In contrast, using two-dimensional lensing distortion pattern,
we are able to model lensing signals from all halos as well as
the projection effect. 
Therefore, we conducted multi-components fitting of the two-dimensional
lensing distortion pattern.
Here, we briefly summarize the fitting method. Full descriptions will be
presented in the forthcoming paper (Okabe et al., in preparation).
We first pixelized the shear distortion pattern into a regular grid
of $2\farcm\times2\farcm$ without any spatial smoothing procedure, 
considering the limited number of the available background galaxies.
The pixelized distortion signals, $\langle{g_\alpha}\rangle$, and the
mean position, $\langle \theta_\alpha\rangle$, in the $n$-th grid are calculated 
with statistical weight, $u_i$,
\begin{eqnarray}
\langle{g_{\alpha}}\rangle &=&
 \frac{\sum\limits_{i\in n} u_{i} g_{\alpha,i}}{\sum\limits_{i\in n}
 u_{i}},  \label{eq:smshear1} \\
\langle{\theta_\alpha}\rangle &=&
 \frac{\sum\limits_{i\in n} u_{i} \theta_{\alpha,i}}{\sum\limits_{i\in n} u_{i}},
 \label{eq:smshear2} \\
u_{g,i}&=&\frac{1}{\sigma_{g,i}^2+{\bar \sigma}_{g}^2}, \label{eq:ug}
\end{eqnarray}
where  $g_{\alpha,i}$ and $\theta_{\alpha,i}$ are the reduced shear and
position of the $i$-th galaxy in the $n$-th grid, respectively.
$\sigma_{g,i}$ is the rms error for shear measurement,
and ${\bar \sigma}_g=(\sum \sigma_{g,i}^2/N)^{1/2}\simeq0.48$ is the
softening constant variance \citep{Okab08,Okab10b}.
Note that this pixelization method is different from that of making
mass maps such as figure \ref{fig1}, where we pixelized the shear
distortion pattern with a Gaussian smoothing kernel (FWHM$=1\farcm67$)
using a shear catalog without the color selection \citep{Okab10b}.
We then conducted the $\chi^2$ fitting given by
\begin{eqnarray}
\chi^2=\sum\limits_{\alpha,\beta=1}^{2}\sum\limits_{n}^{N_{\rm pixel}}
 (g_\alpha(\mbox{\boldmath $\theta$}_n)-g_\alpha(\mbox{\boldmath $\theta$}_n;\mbox{\boldmath
 $p$})^{(\rm model)}) C_{\alpha\beta}^{-1}(\mbox{\boldmath $\theta$}_n) (g_\beta(\mbox{\boldmath
 $\theta$}_n)-g_\beta(\mbox{\boldmath $\theta$}_n;\mbox{\boldmath $p$})^{(\rm model)}),
\end{eqnarray}
where $\mbox{\boldmath $p$}$ is the parameter vector.
$C_{\alpha\beta}$ is the error covariant matrix of shape measurements in the form of
$C_{\alpha\beta}(\mbox{\boldmath $\theta$}_n)=\delta^{\rm
K}_{\alpha\beta}\sigma_g^2(\mbox{\boldmath $\theta$}_n)$, where
$\delta^{\rm K}_{\alpha\beta}$ and $\sigma_g^2(\mbox{\boldmath
$\theta$}_n)$ are a Kronecker delta function and the statistical error of
the pixelized shear \citep{Ogur10}, respectively.

We considered all mass components of C, N, NE, NW, SE and A664.
We adopted an NFW mass profile \citep{nav96,nav97} for each halo model.
The NFW profile can be described with the two
parameters; the virial mass $M_{\rm vir}$ and concentration parameter $c_{\rm vir}$. 
According to cosmological $N$-body simulations based on the CDM universe, it is
well-known that the concentration parameter is a weakly decreasing function of the virial
mass $M_{\rm vir}$ and redshift $z$, e.g., $c_{\rm vir}=7.85\left(M_{\rm
vir}/2\times10^{12}M_\odot\right)^{-0.081}(1+z)^{-0.71}$ 
(Duffy et al. 2008). We here assumed this relation for halos of N, NE, NW, SE and
A664. This is because their angular sizes are likely too small to constrain 
the concentration parameters well, taking into account the fact that
they are at higher redshift ($z\sim0.4$) or less massive objects. 
On the other hand, we treated the concentration parameter as a free parameter for the nearby main
cluster, C.
In addition, the positions of the halo's center ($x_c, y_c$) for C, N, NE, and NW were also parameterized 
so as to avoid a mis-center in mass measurements. 
The center positions for the other components were fixed at those of the brightest galaxy appeared 
in the optical image. As a result, we used $15$ parameters to model the
two-dimensional weak-lensing analysis data. 
The modeling shear distortion pattern, $g_\alpha(\mbox{\boldmath
$\theta$}_n;\mbox{\boldmath $p$})^{(\rm model)}$, is described by a summation of
shear signals of each mass component.
We adopted the Markov Chain Monte Carlo (MCMC) method with standard Metropolis-Hastings sampling.
We restricted the sampling range to $M_{\rm
vir}\leq5\times10^{15}h^{-1}M_\odot$, $c_{\rm vir}\le 50$,
$|x_c-x_{\rm peak}|<1\farcm$ and $|y_c-y_{\rm peak}|<1\farcm$, where $x_{\rm
peak}$ and $y_{\rm peak}$ are the peak coordinates appeared in
weak-lensing mass map.

The obtained virial masses $M_{\rm vir}$ and $M_{500}$, mass within a
sphere of the radius $r_{500}$ are shown in table \ref{tab:mass},
where $r_{500}$ is the radius within which the mean density is 500 times
of the critical density of the universe.
The parameters of NW and SE halos were not constrained well by this analysis, 
because their lensing signals are much smaller than those of the other massive objects. 
The mode of the posterior distribution for the SE halo's virial mass is $M_{\rm
vir}\sim6\times10^{13}h^{-1}M_\odot$. This is typical to galaxy groups rather than clusters,
which implies that it would be mainly due to the lensing signal from the mass
structure associated with the galaxy at $z=0.103$ and possible
background galaxies. 
To investigate this possible projection effect in the SE clump region,
we calculated number densities of galaxies in a given photometric
redshift slice with SDSS photometric data in the same way of \citet{Kawa10},
taking account of an uncertainty of the photometric redshifts.
We calculated peak significance values based on the mean density in the Subaru's field-of-view, 
in order to avoid redshift evolution of the number density of galaxies
and physical scale difference.
We found a $\sim3.5\sigma$ peak at $z=0.1$ with a slice of 
$\delta z=(500~{\rm kms}^{-1})/(1+z)/c\simeq0.002$.
We also found another $\sim3\sigma$ peak in the redshift range of $|z-0.4|\le \delta z=0.01$,  
which corresponds to the velocity dispersion of $\sim2000~{\rm kms}^{-1}$.
Therefore, SE clump is likely to be a superposition of mass structures
at these two different redshifts.
Since a part of the NW halo overlaps with the main cluster C within $r_{500}$, 
the NW halo might be a tidally-truncated substructure.
However, we cannot rule out a possibility that all mass halos are isolated objects because 
we cannot distinguish structures along the line-of-sight in a few tens of Mpc scale observationally.
\begin{table}
  \caption{ The best-fit masses ($M_{\rm vir}$ and $M_{500}$) for clump candidates 
            in the unit of $10^{14}h^{-1}M_\odot$.} \label{tab:mass}
\begin{center}
\begin{tabular}{l||cccccc}
\hline
\hline
Halo Name  & C
        & N 
        & NE
        & NW
        & SE
        & A664 \\ 
\hline
$M_{\rm vir}$  &  $3.38_{-1.47}^{+0.90}$
& $5.82_{-3.07}^{+2.24}$
       &  $2.31_{-1.98}^{+1.19}$
       &  $<0.43$
       &  $<1.05$
       &  $8.12_{-3.29}^{+2.72}$\\
$M_{500}$  &  $2.40_{-1.04}^{+0.64}$
       & $3.21_{-1.66}^{+1.19}$
       &  $1.31_{-1.11}^{+0.65}$
       &  $<0.25$
       &  $<0.58$
       &  $4.32_{-1.75}^{+1.36}$\\
\hline
\end{tabular}
\end{center}
\end{table}

\begin{figure}
  \begin{center}
    \FigureFile(80mm,80mm){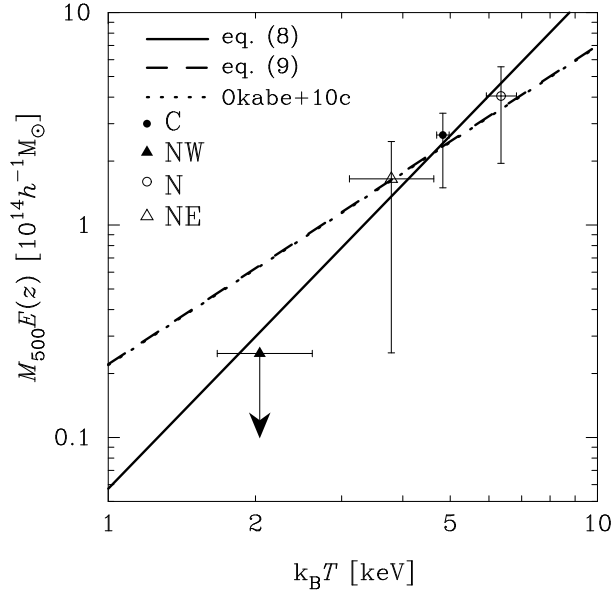}
  \end{center}
  \caption{Mass - temperature relation for the halos. The errors
           are at the one-sigma level. The best-fit power-law functions
 (equations (\ref{eq:m500kt}) and (\ref{eq:m500kt_fixed})) 
           are also overlaid with solid and dashed lines, respectively.
 The dotted lines denote the result of \citet{Okab10c} for a sample of 12 clusters
 based on a joint weak-lensing and X-ray analysis, which almost overlap with the dashed lines.}
  \label{fig13}
\end{figure}

Figure \ref{fig13} shows the $M-k_BT$ relation for the halos.
Again, the errors are at the one-sigma level. 
It should be noted that the mass used here is obtained from the weak-lensing data
alone, which is perfectly independent of the X-ray results and free from
the assumption about the dynamical status of the system.
First, we fit the data into a power-law function setting both the index and normalization 
as a free parameter. The best fit power-law function is
\begin{eqnarray}
   M_{500} E(z)  =  2.63_{-0.63}^{+0.81}
    \left(\frac{k_{\rm B}T}{5~{\rm
     keV}}\right)^{2.38_{-0.95}^{+0.78}}\times 10^{14}h^{-1}M_{\odot},
  \label{eq:m500kt}
\end{eqnarray}
which is plotted as solid lines in figure \ref{fig13}.
Here, we do not take into account intrinsic scatter of the scaling relation (\cite{Okab10c}).
This result is consistent within large errors with what a self-similar model
predicts, $M E(z) \propto T^{1.5}$ \citep{Kais86}. Next, we fit the data into a power-law function 
with the index fixed to that of the self-similar model.
The resultant function is
\begin{eqnarray}
 M_{500} E(z)  =  2.47_{-0.57}^{+0.69}
    \left(\frac{k_{\rm B}T}{5~{\rm keV}}\right)^{3/2}\times 10^{14}h^{-1}M_{\odot},
  \label{eq:m500kt_fixed}
\end{eqnarray}
which is also plotted as dashed lines in figure \ref{fig13}.
Interestingly, the obtained normalization is consistent with that of the previous study 
($2.45^{+0.27}_{-0.24}$;\cite{Okab10c}) 
for a sample of 12 clusters based on weak-lensing and X-ray datasets, which is plotted as dotted lines
in figure \ref{fig13}, though they almost overlap with the dashed lines.
Although we assume the mass-concentration relation for three halos, 
note that the masses at the intermediate scale such as $r_{500}$ hardly depend on the assumed relation.
As mentioned in section \ref{S:gas}, the possible presence of cool cores
and/or AGNs has very limited impacts on the temperature measurement.
We would like to emphasize that this is the first confirmation of a self-similar
scaling relation of mass and temperature for halos discovered by
weak-lensing signals, whereas samples of X-ray selected clusters were used in previous studies of the mass
scaling relation. 
The NW halo's mass expected from its temperature via equation (\ref{eq:m500kt_fixed}) is twice as large as
its actual upper limit, which might indicate that the NW halo is tidally truncated by the primary halo, 
ZwCl0823.2+0425. Another possibility is the probable overestimation of the temperature
because of the contamination as mentioned in section \ref{s:IS}.

Halo detection criteria in weak lensing survey might be affected by the Gaussian 
smoothing scale adopted in mass reconstruction.
We therefore investigated dependence of significance level of peaks on smoothing scales. 
Here, we do not make a selection of background galaxies, as a usual way.
The significance levels changed by only $\le15\%$ as the smoothing scale
changed by $\sim30\%$, and thus we conclude that this effect is negligible for halo detections.

In order to compared the significant level of peaks between the results
with and without the color selection, we made the projected mass density map 
for the best-fit mass model with the same smoothing kernel as figure \ref{fig1}.
This is because the number density of the secure background galaxies is too small to
reconstruct the mass map with the smoothing scale same as figure \ref{fig1}.
The peak $\kappa$ for C and NE halos in the model map are higher than by $2.6\sigma$ and $1.5\sigma$ level of the
reconstruction errors ($\delta\kappa\simeq0.022$) of figure \ref{fig1}.
This implies that a contamination of the member galaxies in the 
shear catalog without color selection (known as a dilution effect) is
significant for C and NE halos.
The NW peak in the model map is about $1\sigma$ level lower, while
SE and N halo peaks are comparable between two maps.
Although we do not have a definite idea why the significance of NW halo decreases,
it is possible that the background selection in the color-magnitude plane did not work well
for some reasons. To make more secular selection of 
background galaxies, we need more detailed multi-band imaging data, as demonstrated by \citet{Umet10}.   
We also computed the contribution for the lensing signal from each halo 
to investigate the projection effect.
The lensing effect from the other halos on C and N halo regions is only about $10\%$. 
In contrast, the contribution from the others accounts for $\sim20\%$, $\sim20\%$ and $\sim40\%$ 
for SE, NE and NW halo regions, respectively.  
As for the observed lensing signals of these three halo regions, therefore, the contribution from the others 
is not negligible.
In particular, the projection effect and/or contribution from nearby halos are critical for 
halo detection with the weak-lensing technique, because they boost the peak strength in the mass map.
For instance, the peak projected density of NW halo alone is only at $\sim2\sigma$
level of the reconstruction error of figure \ref{fig1}. 
This indicates that we need further careful lensing studies for halo
candidates discovered by lensing maps. 
Otherwise, we cannot rule out possibility that 
shear-selected catalog of clusters/groups is contaminated
with some less massive objects.
As for cluster surveys with weak lensing technique alone, this
difficulty should be kept in mind to achieve a complete sample.

\section{Conclusions}\label{s:C}
We observed the field around the galaxy cluster ZwCl0823.2+0425, where several dark matter halos were
found with the weak gravitational lensing technique. 
X-ray emission associated with most of these halos was clearly detected. 
From spectral analysis of the XIS data, we determined temperature for the all halos detected in X-ray, 
and metal abundance for some of them, for the first time. 
With these results, we obtained the X-ray luminosity-temperature 
relation, which is consistent with the former typical results. The gravitational mass of each halo
was obtained from the weak lensing data, which is absolutely independent from the X-ray data
and free from the assumptions about the dynamical status of the systems.
With the lensing and X-ray results, we investigated the mass-temperature relation for the halos, 
which is found to be consistent with the prediction of a simple self-similar model
and previous studies. We first found the self-similar scaling relation of
temperature and mass for halos detected by weak-lensing signals.
Therefore, our study demonstrates the importance of X-ray follow-up
observations of shear-selected clusters, and shows that a joint X-ray and lensing analysis 
is crucial for clusters discovered by the forthcoming weak-lensing surveys, 
such as the one planned with Subaru/Hyper-Suprime-Cam.
It is expected that the weak-lensing surveys will provide us with the uniform catalog of
galaxy clusters irrespective of cluster dynamical states, though there
are still some difficulties as discussed in section \ref{s:D}. 
This is because the lensing technique depends on the cluster masses alone, which is
free from the dynamical state. In contrast, X-ray and SZE observables are obviously 
affected by cluster merger phenomena \citep{Rick01, Taki10}.
A study of mass-scaling relations with statistically complete samples of galaxy clusters 
will enable us to understand cluster baryonic evolution, such as radiative cooling, pre-heating, 
and cluster mergers \citep{Stan10}, and to make a robust mass-proxy 
based on principal component analysis \citep{Okab10c}.
This will be an important step to constrain the equation of state for the dark energy through 
the cluster mass function. The future X-ray projects, such as {\it Astro-H} and {\it eROSITA}, 
will be powerful in this regard.

\bigskip
The authors would like to thank T. Ohashi, K. Sato, and S. Shibata for helpful comments.
We are also grateful to the Suzaku operations team for their support
in planning and executing this observation.
MT and NO were supported in part by a Grant-in-Aid from the
Ministry of Education, Science, Sports, and Culture of Japan
(MT:19740096; NO: 20740099). 
This work is in part supported by a Grant-in-Aid for the COE Program
``Exploring New Science by Bridging Particle-Matter Hierarchy'' and
G-COE Program ``Weaving Science Web beyond Particle-Matter Hierarchy''
in Tohoku University, and for Science Research in a Priority Area "Probing the Dark
Energy through an Extremely Wide and Deep Survey with Subaru
Telescope" (18072001) funded by the Ministry of Education, Science,
Sports and Culture of Japan.  
JET was supported by a Discovery Grant from NSERC Canada.


\end{document}